\documentclass[runningheads,a4paper]{llncs}

\usepackage{amssymb}
\usepackage{graphicx}
\usepackage{multirow}
\usepackage{colortbl}
\usepackage{caption, booktabs}
\usepackage{subcaption}
\usepackage{listings}
\usepackage{xcolor}
\usepackage{algorithm2e}
\usepackage{hyperref}
\usepackage[para]{footmisc}

\usepackage{tikz}
\usepackage{pgfplots}
\usepackage{pgf-pie}
\usepackage{pgfplotstable}
\usepgfplotslibrary{statistics}
\usepackage{float}
\usepackage{adjustbox}

\usepackage{url}
\urldef{\mailsa}\path|{alfred.hofmann, ursula.barth, ingrid.haas, frank.holzwarth,|
\urldef{\mailsb}\path|anna.kramer, leonie.kunz, christine.reiss, nicole.sator,|
\urldef{\mailsc}\path|erika.siebert-cole, peter.strasser, lncs}@springer.com|    
\newcommand{\keywords}[1]{\par\addvspace\baselineskip
\noindent\keywordname\enspace\ignorespaces#1}

\usepackage[printonlyused, nohyperlinks, nolist]{acronym}
\usepackage{mathptmx}                  
\begin{acronym}
 	\acro{fb}[FB]{Full-Body}
	\acro{hh}[HH]{Head-Hand}
	\acro{rl}[RL]{Real-Life}
	\acro{eeg}[EEG]{Electroencephalography}
	\acro{hmd}[HMD]{Head Mounted Display}
	\acro{ve}[VE]{Virtual Environment}
	\acro{vr}[VR]{Virtual Reality}
	\acro{plv}[PLV]{Phase-Locking Value}
	\acro{ccorr}[CCorr]{Circular Correlation Coefficient}
	\acro{rq}[RQ]{Research Question}
	\acro{rrq}[RRQ]{Review Research Question}
	\acro{lsl}[LSL]{Lab Streaming Layer}
	\acro{asr}[ASR]{Artifact Subspace Reconstruction}
	\acro{ik}[IK]{Inverse Kinematics}
	\acro{hypyp}[HyPyP]{Hyperscanning Python Pipeline}
	\acro{ar}[AR]{Augmented Reality}
	\acro{fdr}[FDR]{False Discovery Rate}
	\acro{meg}[MEG]{Magnetoencephalography}
	\acro{fmri}[fMRI]{functional Magnetic Resonance Imaging}
	\acro{fnirs}[fNIRS]{functional Near-Infrared Spectroscopy}
	\acro{fov}[FOV]{Field of View}
	\acro{ui}[UI]{User Interface}
	\acro{ik}[IK]{Inverse Kinematics}
	\acro{llapi}[LLAPI]{Low Level Application Programming Interface}
	\acro{hlapi}[HLAPI]{High Level Application Programming Interface}
	\acro{pun}[PUN2]{Photon Unity Networking 2}
    
\end{acronym} 

\begin{document}


\definecolor{forestgreen}{RGB}{34,139,34}
\definecolor{orangered}{RGB}{239,134,64}
\definecolor{darkblue}{rgb}{0.0,0.0,0.6}
\definecolor{gray}{rgb}{0.4,0.4,0.4}
\definecolor{codegreen}{rgb}{0,0.6,0}
\definecolor{codegray}{rgb}{0.5,0.5,0.5}
\definecolor{codepurple}{rgb}{0.58,0,0.82}
\definecolor{backcolour}{rgb}{0.95,0.95,0.92}

\lstdefinestyle{XML} {
    language=XML,
    extendedchars=true, 
    breaklines=true,
    breakatwhitespace=true,
    emph={},
    emphstyle=\color{red},
    basicstyle=\ttfamily,
    columns=fullflexible,
    commentstyle=\color{gray}\upshape,
    morestring=[b]",
    morecomment=[s]{<?}{?>},
    morecomment=[s][\color{forestgreen}]{<!--}{-->},
    keywordstyle=\color{orangered},
    stringstyle=\ttfamily\color{black}\normalfont,
    tagstyle=\color{darkblue}\bf,
    morekeywords={attribute,xmlns,version,type,release},
    otherkeywords={attribute=, xmlns=},
}

\lstdefinelanguage{PDDL}
{
  sensitive=false,    
  morecomment=[l]{;}, 
  alsoletter={:,-},   
  morekeywords={
    define,domain,problem,not,and,or,when,forall,exists,either,
    :domain,:requirements,:types,:objects,:constants,
    :predicates,:action,:parameters,:precondition,:effect,
    :fluents,:primary-effect,:side-effect,:init,:goal,
    :strips,:adl,:equality,:typing,:conditional-effects,
    :negative-preconditions,:disjunctive-preconditions,
    :existential-preconditions,:universal-preconditions,:quantified-preconditions,
    :functions,assign,increase,decrease,scale-up,scale-down,
    :metric,minimize,maximize,
    :durative-actions,:duration-inequalities,:continuous-effects,
    :durative-action,:duration,:condition
  }
}

\lstdefinelanguage{Srv}{
keywords = [1]{bool, uint8, int32, uint64, float32, float64, string, Header, Point, Quaternion, time},
comment=[l]{\#}
}

\lstdefinestyle{mystyle}{
  backgroundcolor=\color{backcolour},   commentstyle=\color{codegray},
  keywordstyle=\color{codegreen},
  numberstyle=\tiny\color{codegray},
  stringstyle=\color{codepurple},
  basicstyle=\ttfamily\footnotesize,
  breakatwhitespace=false,         
  breaklines=true,                 
  captionpos=b,                    
  keepspaces=true,                 
  numbers=left,                    
  numbersep=5pt,                  
  showspaces=false,                
  showstringspaces=false,
  showtabs=false,                  
  tabsize=2
}
\lstset{style=mystyle}

\mainmatter  

\title{Effects of Human Avatar Representation in Virtual Reality on Inter-Brain Connections}

\titlerunning{Effects of Human Avatar Representation in VR on Inter-Brain Connections}

\author{Enes Yigitbas \and Christian Kaltschmidt}

\authorrunning{Enes Yigitbas et al.}

\institute{Paderborn University\\ Zukunftsmeile 2, 33102 Paderborn, Germany\\
\email{firstname.lastname@upb.de}, 
}

%
%

\toctitle{Lecture Notes in Computer Science}
\tocauthor{Authors' Instructions}
\maketitle

\begin{abstract}
Increasing advances in affordable consumer hardware and accessible software frameworks are now bringing Virtual Reality (VR) to the masses. Especially collaborative VR applications where different people can work together are gaining momentum. In this context, human avatars and their representations are a crucial aspect of collaborative VR applications as they represent a digital twin of the end-users and determine how one is perceived in a virtual environment. When it comes to the effect of avatar representation on the end-users of collaborative VR applications, so far mostly questionnaires have been used to assess the quality of avatar representations. A promising alternative to objectively measure the effect of avatar representation is the investigation of inter-brain connections during the usage of a collaborative VR application. However, the combination of immersive VR applications and inter-brain connections has not been fully researched yet. Thus, our work investigates how different human avatar representations (real (RL), full-body (FB), and head-hand (HH)) affect inter-brain connections. For this purpose, we have designed and conducted a hyperscanning study with eight pairs. The main results of our hyperscanning study show that the number of significant sensor pairs was the highest in the RL, medium in the FB, and lowest in the HH condition indicating that an avatar that looks more like a real human enables more significant sensor pairs to appear in an EEG analysis.

\end{abstract}

\section{Introduction}\label{sec:intro}
Nowadays, VR applications are used in various application domains such as education, medicine, tourism, entertainment, etc. Especially collaborative VR applications where different people can work together are gaining momentum. In this context, human avatars and their representations are a crucial aspect of collaborative VR applications as they represent a digital twin of the end-users and determine how one is perceived in a virtual environment.  

Zanbaka et al. found out that not only the presence of real humans but also the presence of virtual humans increases task performance \cite{social_presence_performance}.
This is a promising finding that working collaboratively in \ac{ar} or \ac{vr} environments can boost the working outcome. When interacting with other users in VR, the concept of social presence, which describes the feeling that the other participant is present in the \ac{ve} and can be interacted with, can be observed \cite{social_presence}. In collaborative applications, a higher degree of social presence is desired \cite{social_presence_is_desired}.

Not only the aspect of presence but also the aspect of avatar representation is important for accomplishing tasks in collaborative VR applications. Several approaches of avatar representations have an impact on the user experience of a \ac{vr} or \ac{ar} application \cite{self_avatars,mini-me,influence_avatar_representation,ar_social_presence}.

When it comes to the effect of avatar representation on the end-users of collaborative VR applications, so far mostly questionnaires have been used to assess the quality of avatar representations. As the answers to a questionnaire are a subjective measure, it is difficult to derive design recommendations for the representation of avatars in different usage contexts of collaborative VR applications. A promising alternative to objectively measure the effect of avatar representation is the investigation of inter-brain connections during the usage of a collaborative VR application. 

An interesting phenomenon that can be observed during social interaction between two or more people is that brain areas of the interacting people can connect by synchronization \cite{brain-to-brain_social,inter_brain_social}.
To find brain areas of two or more interacting people that are in synchrony, their brain activity has to be measured. Typically, this is being done with \ac{eeg}, \ac{meg}, \ac{fmri}, or \ac{fnirs} devices.
Measuring brain activity from multiple people at the same time is called hyperscanning. With several mathematical methods, it can be calculated if connections, namely inter-brain connections, between the subjects' brains were able to establish \cite{ccorr}. 
Szymanski et al. found out that the degree of synchronization of two people correlates to their task performance in a given task \cite{hyperscanning_better_performance}.
This finding suggests that a higher amount of inter-brain connections between two users during a cooperative task is desired.

Since the first hyperscanning study \cite{hyperscanning_first_study} was conducted, it is known that brain areas of interacting people can synchronize. These connections can also be observed when people interact in a virtual environment \cite{hyperscanning_vr}.
Yet, only two studies investigated inter-brain connections while using \acp{hmd} \cite{hyperscanning_vr,gaze_vr_hyperscanning}.
Thus, the combination of immersive VR applications and inter-brain connections in general is not researched well even though the degree of synchrony between the participants is expected to correlate with the performance during tasks \cite{hyperscanning_better_performance}.
In \cite{gaze_vr_hyperscanning}, Gumilar et al. investigated how the eye gaze of a virtual avatar affects inter-brain connections between participants. Our work focuses on the avatar representation in a \ac{ve}.
That leads to the \ac{rq}:
\begin{quote}
	\textbf{\ac{rq}:} How do different human avatar representations affect inter-brain connections?
\end{quote}

To answer this \ac{rq}, an investigation of different human avatar representations in \ac{vr} is combined with a hyperscanning measurement.

 For this purpose, we have conducted a hyperscanning study with eight pairs. Based on a simple motoric task where the participants had to navigate a ball through a maze, we analyzed different conditions and their implications on inter-brain connectivity. The first condition was done in real life only, so the participants navigated the ball through a custom-made real-physical maze. The other two conditions took place in a \ac{ve}. A room in the \ac{ve} was designed similar to the room the \ac{rl} condition took place in.
Also, the virtual maze was designed exactly like the maze in real life. In the two \ac{vr} conditions, the participants saw each other as a \ac{fb} avatar as well as an \ac{hh} avatar.
During all three conditions, \ac{eeg} data from both participants was measured simultaneously. 

The main results of our hyperscanning study show that the number of significant sensor pairs was the highest in the \ac{rl} condition, medium in the \ac{fb} VR condition, and lowest in the \ac{hh} condition indicating that an avatar that looks more like a real human enables more significant sensor pairs to appear. Furthermore, when looking at specific sensor pairs that were significant in the majority of participant pairs, the \ac{fb} condition stands out as it shows six inter-brain connections as opposed to only one inter-brain connection during the \ac{hh} condition. However, the results are a first indicator and further steps are necessary to conclude generalizable findings.

The rest of the paper is structured as follows: Section \ref{sec:related_work} presents related work about human avatar representations and similar hyperscanning studies. Section \ref{sec:conceptual_solution} describes the conception and implementation of the system environment to conduct the motivated hyperscanning study. In Section \ref{sec:usability_evaluation}, we present the results of the study and finally, in Section \ref{sec:conclusion}, we conclude the paper and give an outlook for potential future work.

\section{Related Work}\label{sec:related_work}

In this section, we draw on prior research dealing with the topics of human avatar representations and investigation of different human avatars with a hyperscanning approach.

\subsection{Human Avatar Representation}

Heidicker et al. \cite{influence_avatar_social_vr} investigated whether different avatars have an impact on the communication between the participants during a collaborative task in \ac{vr}. For the experiment, the participants had to do the \textit{Desert Survival Task} which requires the participants to rank several items regarding their importance for surviving in the desert \cite{desert_survival}.
After the experiment, the participants had to fill out three questionnaires namely the \textit{Networked Mind Social Presence Questionnaire} \cite{networked_mind}, \textit{Slater Usoh Steed Presence Questionnaire (SUS)} \cite{sus} and the \textit{NASA TLX Questionnaire} \cite{nasa_tlx}. The results of those questionnaires revealed that the \textit{NASA TLX} did not find any significant difference between the avatar representations. Yet, some subcategories of the \textit{Networked Mind Social Presence Questionnaire} showed significant differences. The one-to-one mapped full-body avatar and the head-hand avatar show increased \textit{co-presence} scores in contrast to the idle full-body avatar.
Likewise, in the subcategory \textit{behavioral interdependence}, the \textit{idle full-body} avatar achieved significantly lower scores than the other avatars. The \textit{SUS} questionnaire revealed that the one-to-one mapped avatar shows a significantly higher presence score than the other two avatars.

Yoon et al. \cite{ar_social_presence} investigated the effect of avatar appearance on social presence in an augmented reality collaboration. In this study, six different avatars were investigated in two experiments. They were divided into three body types and two art styles. The art styles were a realistically looking 3D-model and a 3D-model in cartoon style. Each of the art styles had the three body types full-body, upper-body, and head-hand. The results of the study showed that the art style of the avatars did not have any effect on social presence.
The model style, on the other side, influences social presence.
The more complete the body of the avatar is, the higher the social presence score of the avatar. However, it is important to note that the difference in that score for the upper-body and full-body avatars showed no significant difference. The authors assume that the viewing angles caused the avatar of the other participant to be not fully visible, thus the two avatars looked too similar to make an impact on the social presence score.

Aseeri et al. \cite{influence_avatar_representation_this} investigated the influence of avatar representation on interpersonal communication in virtual social environments. In their study, they compared three different avatars while doing an experiment procedure that included multiple tasks.
While the first avatar is just a representation of the used \ac{hmd} and its controllers in the \ac{ve}, the other two avatars are full-body representations.
The first full-body avatar is a 3D scan of the participant that was converted to a 3D model for the \ac{ve}.
The second full-body avatar is a live recording of the participant that is streamed into the \ac{ve}.
With these avatars, the participants had to do different different tasks in \ac{vr}. The results showed that the full-body avatar with the video live stream caused the most interpersonal trust compared to the other two avatars.
It also caused the participants to focus more on facial expressions and engage in mutual gaze behavior more than using the other two avatars.
The two full-body avatars evoked higher scores of co-presence, a subcategory of the \textit{Networked Mind Social Presence Questionnaire}.
However, in the condition with the scanned full-body avatar, the participants paid more attention to the body posture of the other participant than in the conditions with the live recording full-body and the \ac{hmd} and controllers-only avatars.

\subsection{Hyperscanning Studies}

In the following, we focus on hyperscanning-related studies that include an experiment with \ac{vr} \acp{hmd}.

Gumilar et al. \cite{hyperscanning_vr} conducted a comparative study on inter-brain synchrony in real and virtual environments using hyperscanning. In their study, they have investigated if an experiment executed in both the real world and a \ac{ve} with \ac{vr} \acp{hmd} yields similar results regarding inter-brain connections between two participants. For this purpose, they replicated the study by Yun et al. \cite{fingerpointing} which used a finger-pointing task.
In both experiments, the participants were seated facing each other and pointing with the index finger of a hand at each other.
One participant was elected the leader and the other participant was the follower.
The leader started to move their finger around and the follower had to follow that finger with their own finger.
In Gumilar et al.'s study, this experiment was conducted in the real world and a \ac{ve}.
A full-body avatar was used to represent the participant in the \ac{ve}.
The measured \ac{eeg} data was statistically analyzed.
To determine how many inter-brain connections were able to be established, \ac{plv} was used. The results revealed that experiments in \ac{vr} can evoke inter-brain connections between participants in a similar way to experiments solely executed in the real world.

In a follow-up study on inter-brain synchrony and eye-gaze direction during collaboration in VR, Gumilar et al. \cite{gaze_vr_hyperscanning}  investigated how the eye gaze of a virtual avatar influences inter-brain connections between two participants. The results show that avatars with a natural eye gaze have a significant impact on inter-brain connections during collaborative tasks in \ac{vr}.
This difference was especially apparent in the alpha band during the natural gaze condition.
However, on average fewer inter-brain connections are established in the natural gaze condition in the alpha band per participant pair but the standard deviation across all participant pairs is higher than for the other conditions.
Yet, the authors argue that social cues in \ac{vr} have a significant impact on inter-brain connections that needs to be investigated further.  

\subsection{Discussion}

Our literature research revealed that there are many works dealing with the topic of avatar representation and appearance as well as their influence on social presence and interpersonal collaboration. However, most approaches are based on subjective measurements relying on questionnaires. The \ac{eeg} hyperscanning-related studies, on the other hand, did not primarily compare avatar representations against each other, or the conducted experiments are based solely on statistical comparisons of established inter-brain connections. To overcome these issues an investigation of different human avatar representations in \ac{vr} is combined with a hyperscanning measurement in our work.

\section{Conception and Implementation}\label{sec:conceptual_solution}

To answer the motivated research question and investigate how different human avatar representations affect inter-brain connections of two people interacting with each other in \ac{vr}, three conditions have been investigated in a hyperscanning study.
The first condition is the task in the \acf{rl} without using any \acp{hmd}. The second and third conditions are executed in \ac{vr} with \acp{hmd}. During the last two conditions, the participants see each other with different avatars, namely as a \acf{fb} and a \acf{hh} avatar. To investigate the effects that these conditions have on the participant's brain activity, an appropriate experiment had to be designed. In the following, we focus on and present the conception and implementation of the system that is required to conduct the hyperscanning study. 

\begin{figure}
	\centering
	\includegraphics[width=\linewidth]{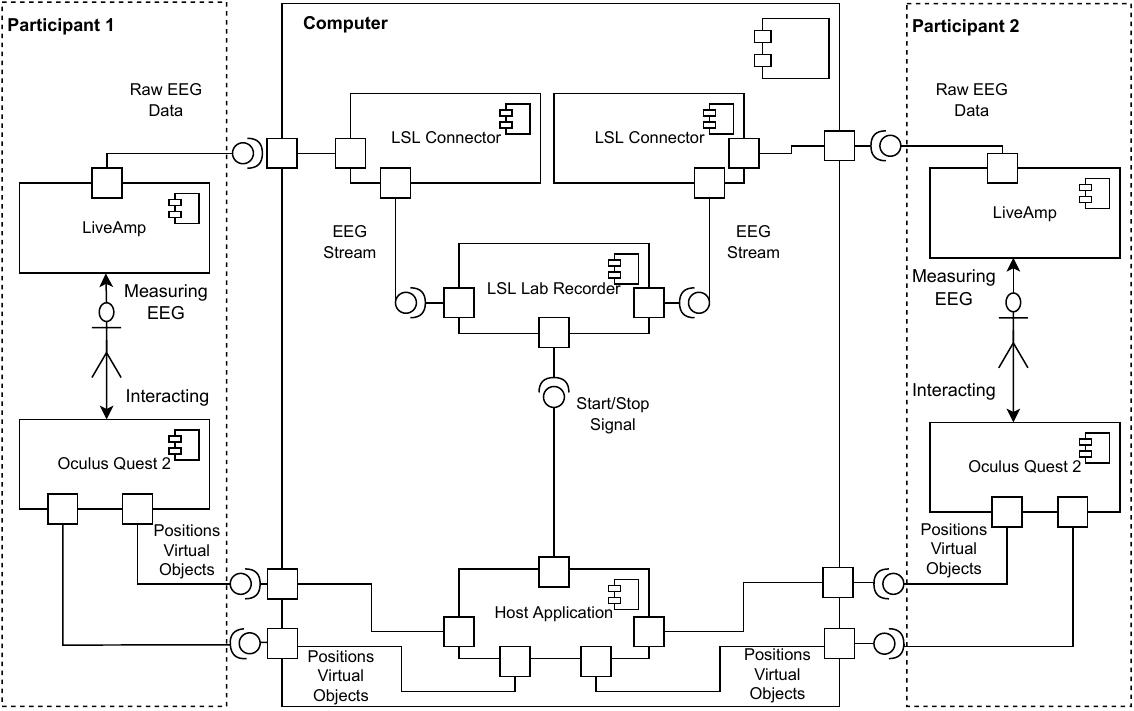}
	\caption{Overview of the hardware and software setup}
	\label{fig:impl_system_overview}
\end{figure}

Figure \ref{fig:impl_system_overview} shows an overview of the whole system including devices and software. The sides of the component diagram show the devices the participants are wearing.
We have decided to use the \ac{hmd} Meta Quest 2 as it can be worn in combination with a \ac{eeg} cap. A \ac{ve} for the experiment is the core element of the client application running on the Quest 2. As two participants are interacting with each other, their application instances must be synchronized.
A host application running on a separate computer receives and distributes the states and positions of the virtual objects to all connected clients. Both \acp{hmd} are connected to the computer and the host application via the network. Additionally to the application, \ac{eeg} data is measured from the participants using an \ac{eeg} measurement device. The LiveAmps of both participants in Figure \ref{fig:impl_system_overview} are connected to the \ac{lsl}\footnote{\url{https://github.com/sccn/labstreaminglayer}} connectors which are pulling the \ac{eeg} data continuously.
These \ac{lsl} connectors are running on the same computer and making the \ac{eeg} data stream accessible to other programs.
The \ac{lsl} Lab Recorder receives the \ac{eeg} data stream from the connectors and saves it to a single file. This data is enriched with time stamps when the experiment is started on the host application by a connector that streams data from the host application to the \ac{lsl} Lab Recorder.

The avatars were created using the program Makehuman\footnote{\url{http://www.makehumancommunity.org/}} version 1.2.0. Makehuman is a program dedicated to creating virtual human avatars. It offers various parameters to adjust the appearance of the avatar.

\section{Evaluation}\label{sec:usability_evaluation}

To investigate the research question if different avatar representations affect inter-brain connections, a hyperscanning study was conducted based on the previously described system design. 

\subsection{Experiment Overview}
\label{subsec: method}

\begin{figure}
	\centering
	\includegraphics[width=\textwidth]{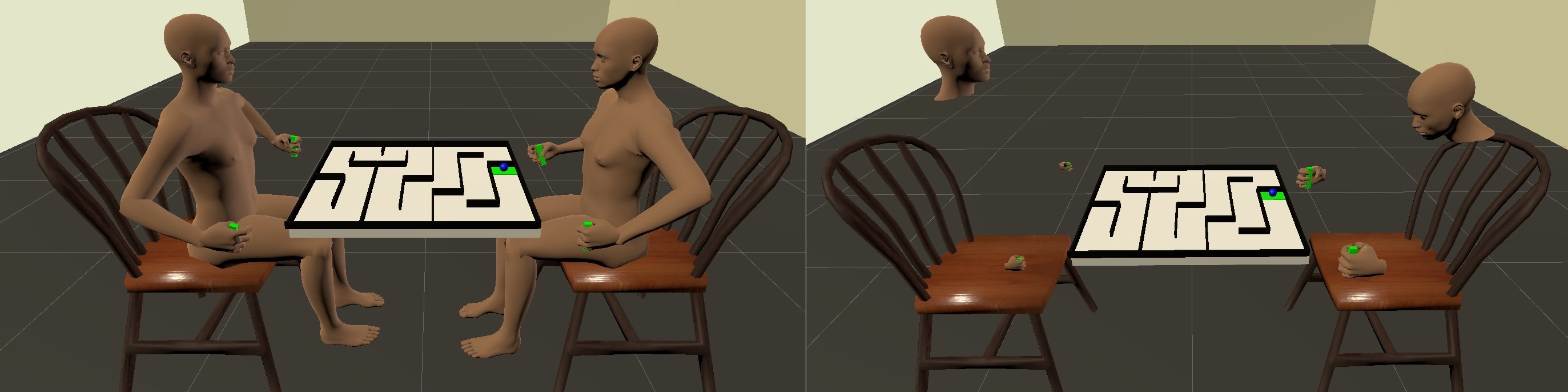}
	\caption{Both avatars during the \ac{vr} conditions. The left side shows the \ac{fb} avatars, and the right side shows the \ac{hh} avatars.}
	\label{fig:eval_both_avatars}
\end{figure}

In this study, two different virtual avatars are compared to each other.
The \ac{fb} avatar presents a complete human body. It is designed to look not too gender-specific as there are studies that suggest differences in inter-brain connections when male and female interact with each other compared to interactions between female and female or male and male \cite{gender_diff1,gender_diff2}. Also, this avatar does not have many details like hair or specific clothes to follow the paradigm of Gumilar et al. in their study \cite{hyperscanning_vr}.
The second avatar for the condition \ac{hh} is an avatar that just has a head and hands. To create this avatar, the full body avatar was imported to Blender and the head and hands were cut off and exported as a distinct model. This approach ensured that both avatars look alike and do not introduce unwanted visual cues. Figure \ref{fig:eval_both_avatars} shows the \ac{fb} avatars on the left side and the \ac{hh} avatars on the right side while they are doing the \ac{vr} conditions.

\subsubsection{Task Design}

\begin{figure}[h!]
	\centering
	\includegraphics[width=0.4\textwidth]{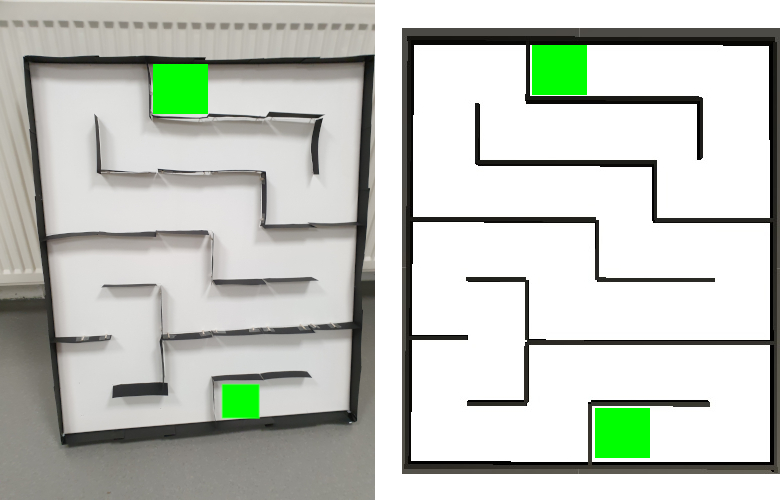}
	\caption{Comparison of maze in real life (left) and \ac{vr} (right). During all conditions, a ball has to be navigated between the green markers continuously back and forth.}
	\label{fig:parkour_comparison_rl_vr}
\end{figure}

Following \cite{hyperscanning_vr} and \cite{no_phyiscal_co_presence}, the task is an easy motoric movement that involves the navigation of a table tennis ball through a maze. The maze is built on a lightweight board. It is important to note, that both the lightweight board and the task consider that the participants do not make too big movements with their upper body because those movements might induce muscle artifacts in the \ac{eeg} signal. Figure \ref{fig:parkour_comparison_rl_vr} shows the board with the maze built on top in real life and the virtual maze next to it. The goal is to navigate a table tennis ball from one target area (green markers in Figure \ref{fig:parkour_comparison_rl_vr}) to the other green-marked target area by tilting the board. If the other target area is reached, the participants have to navigate the ball back to the first area. This is repeated until the timer runs out. 

Each condition lasts four minutes. The experiment in \ac{vr} is designed similarly to the \ac{rl} condition. For the \ac{vr} conditions, a digital twin of the maze was designed to ensure a comparable task. During the \ac{vr} conditions, the participants tilt the board by grabbing virtual handles with the HMD's controllers. The following conditions were tested: \acf{rl}, \acf{fb}, and \acf{hh}.
The \ac{rl} condition has always been the first condition.
It served as a baseline condition. Conditions \ac{fb} and \ac{hh} were randomized across the participant pairs to prevent a temporal bias. 
Each of the conditions lasted four minutes to ensure enough \ac{eeg} data to be produced.

\subsubsection{Study Setup}
\begin{figure}
	\centering
	\includegraphics[width=\textwidth]{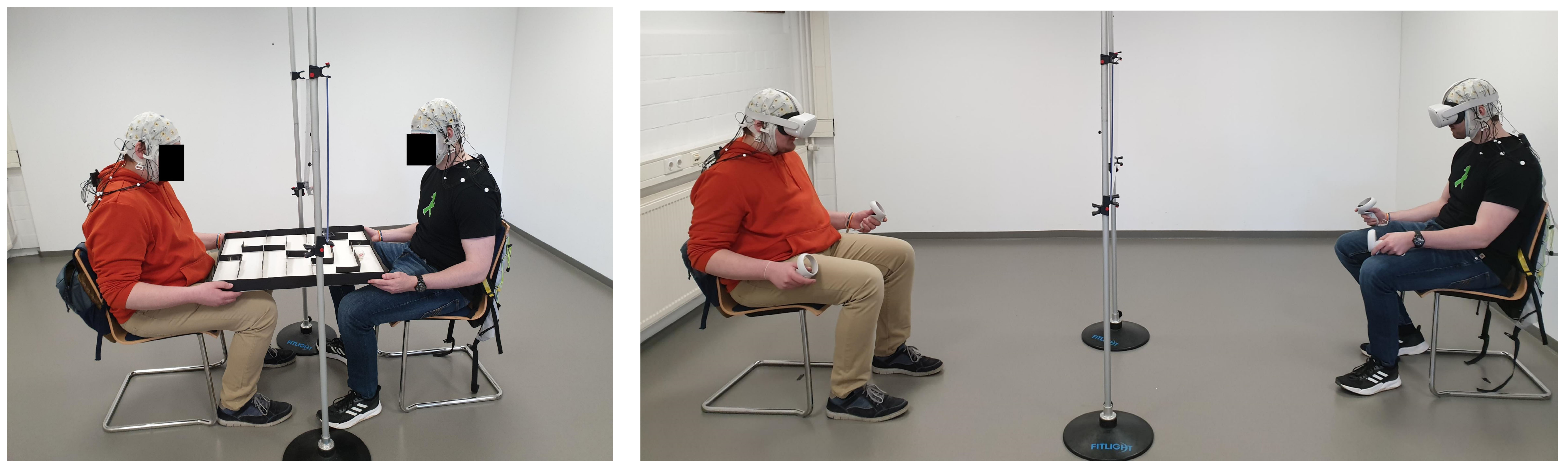}
	\caption{Participants seated during \ac{rl} condition (left) and \ac{vr} conditions (right)}
	\label{fig:positions_rl_condition}
\end{figure}

The experiment took place in the laboratory of the \textit{Exercise- and Neuroscience Department of Paderborn University}. For the experiment execution, the subjects were seated facing each other.
Both held the board as shown in Figure \ref{fig:positions_rl_condition} (left).
During the \ac{vr} conditions, the participants were seated further away from each other. Figure \ref{fig:positions_rl_condition} (right) shows the participants while they are doing the \ac{vr} conditions.
This positioning ensured that the participants did not interfere with each other while they were moving their arms. As the LiveAmp receivers were connected to the computer, they were approximately two meters away from the LiveAmp transmitters, respectively the participants.

\subsubsection{Study Procedure}

The study procedure started with recruiting participants. Upon email invitation, 16 participants could be recruited who were formed into eight groups of pairs. After the participants arrived at the laboratory, they were explained the experiment and why the experiment was being done. Also, the participants were informed that they were allowed to cancel the experiment at any time without any reason or further consequences. After that, the participants filled out a general demographic questionnaire about their age, gender, and experience with \ac{vr} headsets. After the introduction, the \ac{eeg} caps were prepared. As this process is time-consuming, a second person helped with the preparation. First, the circumference of each participant's head was measured with a tape measure. According to the participant's size of their head, a fitting \ac{eeg} cap was chosen and put on the participant's head. To ensure that the caps are placed as similar as possible on every subject's head, the cap was shifted to a specific position so that sensor \textit{Cz} is at half the distance between the nasion and the inion and half the distance between both ears. 
Eventually, a conductivity gel was applied to reduce the impedance between \ac{eeg} sensors and the skin on the head. A blunt syringe was used to inject the gel between the \ac{eeg} sensors and the scalp.
The impedance of each sensor was reduced to $< 50 k\Omega$. When the participants were prepared for the measurement, the server for the \ac{vr} application was started. After all data acquisition devices and applications were started, the participants were seated and the experiment was executed following the above-mentioned descriptions under task design. When all conditions finished, the \ac{vr} headsets and \ac{eeg} caps were pulled off. The whole experiment process took around two hours for each pair.

\subsubsection{Data Preprocessing and Synchronization Metric}

The first step to be able to analyze the data is a preprocessing of the \ac{eeg} data. Following Makoto's preprocessing pipeline\footnote{\url{https://sccn.ucsd.edu/wiki/Makoto's_preprocessing_pipeline}}, the \ac{eeg} data for this study was preprocessed accordingly. For this process, Matlab 2021b\footnote{\url{https://de.mathworks.com/products/matlab.html}} and its widely used plugin EEGLAB v2021.0\footnote{\url{https://sccn.ucsd.edu/eeglab/index.php}} was utilized.

To investigate the \ac{rq}, the \ac{eeg} data has to be analyzed statistically. For this process, a synchronization metric is needed.
Burgess proposes to favor \ac{ccorr} over \ac{plv} when doing a hyperscanning-related connectivity analysis \cite{ccorr}.
Several authors \cite{no_phyiscal_co_presence,gaze_vr_hyperscanning,ref_ccorr1} followed Burgess' suggestion and used \ac{ccorr} as a synchronization metric. In our work, \ac{ccorr} is also used to assess the synchronization between two \ac{eeg} sensors.
The analysis is being done in Python\footnote{\url{https://www.python.org/}} using the packages MNE-Python\footnote{\url{https://mne.tools/stable/index.html}}\cite{mne} and \ac{hypyp}\footnote{\url{https://github.com/ppsp-team/HyPyP}}\cite{hypyp}.\\
First, the preprocessed data of each participant was loaded and split into epochs using MNE-Python. An epoch length of 6s with an overlap of 5s of the previous epoch was chosen based on the results of Fraschini et al. who suggest an epoch length between 4s and 8s \cite{epoch_length}.
Each condition lasted 240s which means overall there were 235 epochs for every condition. The \ac{ccorr} values are computed for each pair of two sensors from the participant pair on each of the 235 epochs on the beta (14Hz - 30Hz) frequency band. We have primarily focused on the beta frequency band as this one is characteristic of normal consciousness and active concentration. 

\subsubsection{Statistical Analysis}

To find out if different human avatars in a \ac{vr} application affect inter-brain connections, a statistical analysis has to be carried out on the \ac{eeg} data. The statistical analysis is applied to the aforementioned \ac{ccorr} values. Similar to Wikström et al. \cite{no_phyiscal_co_presence}, this study compares the \ac{ccorr} values of each sensor pair of a real participant pair to its according sensor pair of a control participant pair as a control measurement.
The control pair consists of one participant from the real pair and one participant from another pair. The idea is that the \ac{ccorr} values of a control participant pair should be purely random, while the \ac{ccorr} values of a real participant pair are not. If the values of a real participant pair differ significantly from the control pair, the investigated sensor pair is considered a significant inter-brain connection.

The significance test depends on the distribution of the data.
To test if the distributions of the sensor pairs are normally distributed, the Shapiro-Wilk-Test was conducted on all epochs of a sensor pair on both the real and control participant pairs.
This test revealed that some of the sensor pairs are normally distributed and some are not. Because the permutation t-test has no pre-conditions on the underlying distribution, it was chosen to determine significant sensor pairs. The permutation t-test tests the null hypothesis that two given data samples are taken from the same population with a confidence interval of 95\%. That means, if a sensor pair has a p-value \textless 0.05, the null hypothesis is rejected and the sensor pair is considered significant. The permutation t-test was executed with 10'000 permutations. Afterward, \ac{fdr} correction was applied.
If the sensor pair is still significant after the \ac{fdr}, that sensor pair is regarded as an inter-brain connection. In this work, significant sensor pairs and inter-brain connections are used synonymously.

\subsection{Results and Analysis}
\label{subsec: results}

In total, 16 participants (12 male and 4 female) could be recruited to participate in our study. The participants were between 22 and 33 years old (mean = 25.875 years). Most of the participants used \ac{vr} applications several times before this study and can be seen as tech- and VR-savvy.

To answer the \ac{rq} \textit{How do different avatar representations affect inter-brain connections?} a statistical analysis as described above has been applied to the measured \ac{eeg} data. It was analyzed how many participant pairs showed a significant sensor pair between sensors $i$ and $j$ with $i,j \in [1, ..., 32]$ and $i$ is a sensor of participant 1 and $j$ is a sensor of participant 2. Table \ref{tab:total_connections_beta} shows significant sensor pairs of the beta band. The rows shows the amount of significant sensor pairs in their respective condition.
The first eight columns indicate in how many participant pairs the significant sensor pairs appeared. In the last column, the total amount of significant sensor pairs is listed.

\begin{table}[h]
	\centering
	\begin{tabular}{r| c c c c c c c c |c}
		
		Appearance in participant pairs & 1& 2& 3& 4& 5& 6& 7& 8& Total\\
		\hline
		\ac{rl} & 393 & 251 & 96 & 15 & 4 & 0 & 0 & 0 & 759\\
		\ac{fb} & 377 & 275 & 85 & 11 & 6 & 0 & 0 & 0 & 754\\
		\ac{hh} & 399 & 247 & 77 & 20 & 1 & 0 & 0 & 0 & 744\\
		
	\end{tabular}
	\caption{Table showing how many significant sensor pairs appeared for a certain amount of participant pairs in the \textbf{beta} band. 
		Rows show the conditions \acf{rl}, \acf{fb}, and \acf{hh}. Columns show the amount of significant connections for a given amount of participant pairs.}
		\label{tab:total_connections_beta}
\end{table}

\begin{figure}[h!]
	\centering
	\includegraphics[width=0.5\textwidth]{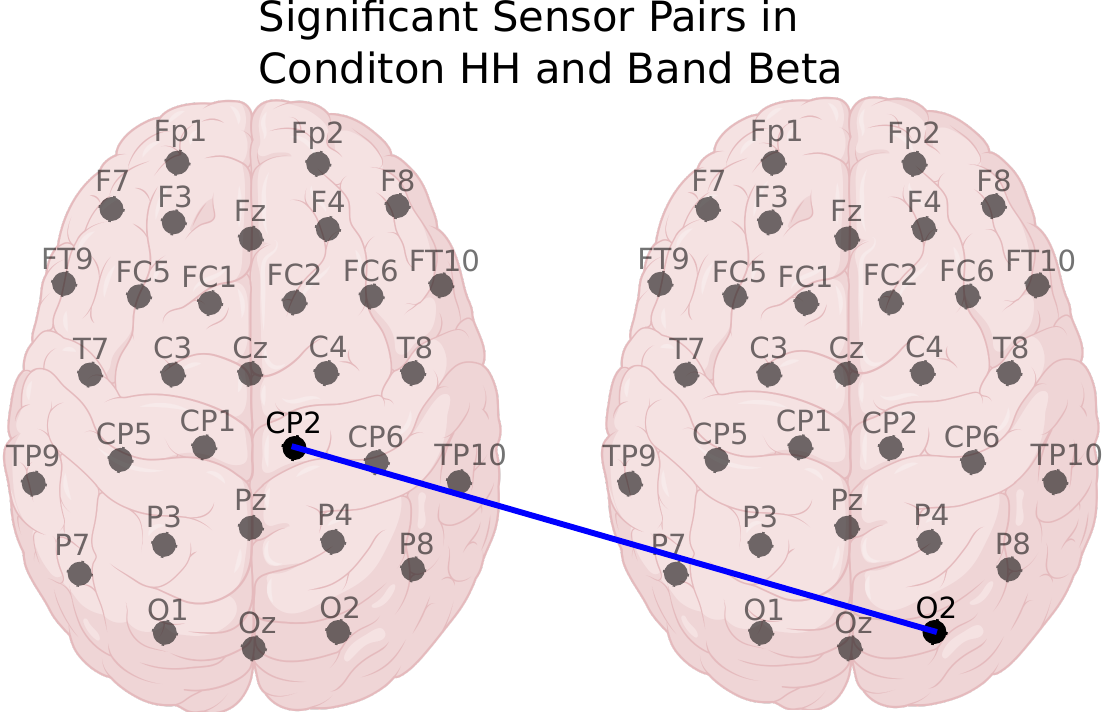}
	\caption{Significant sensor pairs for all conditions in frequency band beta. Blue lines indicate sensor pairs which are significant in five participant pairs.
	}
	\label{fig:connectivityplot_beta}
\end{figure}

Table \ref{tab:total_connections_beta} shows how many significant sensor pairs appeared in the beta band. In total, 759 significant sensor pairs occurred in the \ac{rl} condition, 754 in the \ac{fb} condition, and 744 in the \ac{hh} condition.
In this band, all conditions caused significant sensor pairs to appear on each condition. \ac{rl} condition has four sensor pairs that are significant in five participant pairs, in the \ac{fb} condition six significant sensor pairs appeared and in the \ac{hh} condition one significant sensor pair was found.
Figure \ref{fig:connectivityplot_beta} indicates for the \ac{rl} condition that significant pairs between the sensors FC2 and F7, Oz and F7, P4 and O1 and Cp5 and C4 are established.
Condition \ac{fb} with six significant sensor pairs has pairs between F8 and FC1, FC6 and FP1, CP2 and TP10, O2 and F7, P3 and Cz and TP9 and FP2. The one significant sensor pair in condition \ac{hh} is between O2 and CP2.

\begin{figure}[h!]
	\centering
	\includegraphics[width=0.5\textwidth]{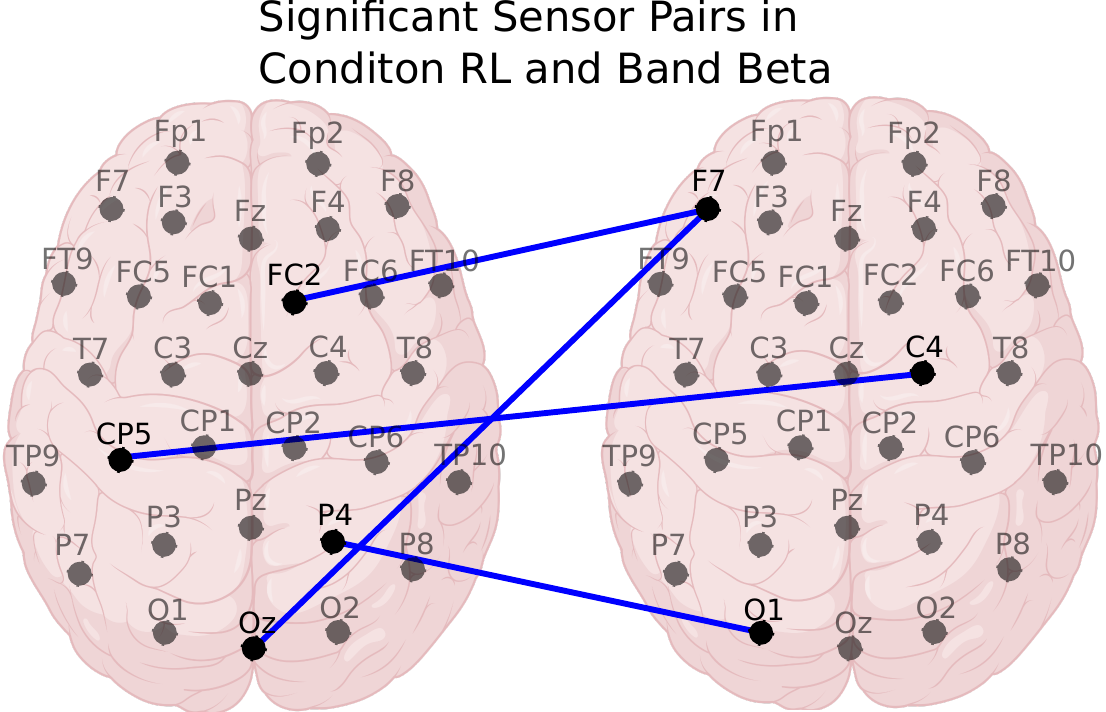}
\end{figure}
\begin{figure}[h!]
	\centering
	\includegraphics[width=0.5\textwidth]{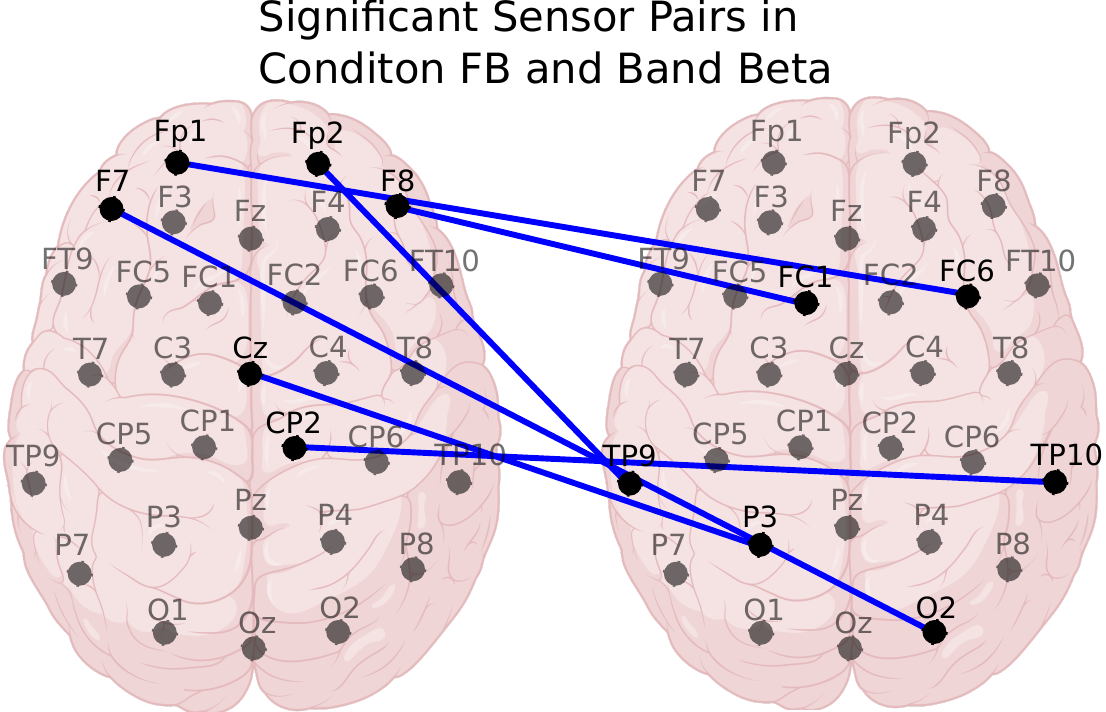}
\end{figure}

\subsection{Discussion and Threats to Validity} 

In the following, we discuss the main results of the specific significant sensor pairs and their potential implications. 
As the condition \ac{rl} is the most naturalistic, we formulate the first hypothesis $H_1$ as:
\begin{quote}
	$\mathbf{H_1:}$ Condition \acf{rl} yields the highest amount of significant sensor pairs.
\end{quote}
Following this assumption, it is expected that an avatar that looks more like a real human enables more significant sensor pairs to appear. Therefore, hypothesis $H_2$ is:
\begin{quote}
	$\mathbf{H_2:}$ Condition \acf{fb} yields a higher amount of significant sensor pairs than condition \acf{hh}.
\end{quote}

Looking at the results of significant sensor pairs in the beta band (see Table \ref{tab:total_connections_beta}), we can conclude that there is a positive trend to accept both hypotheses as condition \ac{rl} has 759, condition \ac{fb} 754 and condition \ac{fb} 744 significant sensor pairs. The total amount of significant sensor pairs in condition \ac{rl} is $0.6\%$ higher than condition \ac{fb} and condition \ac{fb} has $1.3\%$ more connections than condition \ac{hh}. However, it should be noted that the relative differences are quite small, and further studies are required to make conclusions. Therefore, in the following, we discuss the threats to validity that might arise from the design but also the execution of the experiment.

First of all, we had a small number of participant pairs of only eight pairs where only two female pairs participated. Most of the participants have been computer science students mostly familiar with VR technology. A larger user study with heterogeneous participants and pairs is needed to increase the generalizability of the findings. Another threat to validity is the \ac{ve}.
As the physics in \ac{vr} is different than in real life, the expectations of the participants might not always be met which can irritate them. 

Furthermore, the preparation and measurement process of the \ac{eeg} system involves a source for bias. Each human has a differently shaped head. Therefore, it is not possible to fit the \ac{eeg} cap on every participant's head the same.
The \ac{eeg} sensors will always have a different position.
This makes it possible that in some of the experiment appointments, some inter-brain connections were not found because the sensors were not in a position to detect the appropriate signal. Another threat is putting on the \ac{vr} glasses over the \ac{eeg} cap. This procedure potentially misplaces sensors which might have a dramatic effect on the sensor's signal.
Even though, some of the effects like a flat signal or bursts in the \ac{eeg} signal can be detected during the pre-processing step. Another effect that cannot be eliminated completely during the experiment is social cues like laughing or short comments from the participants. Even though the participants were asked beforehand that they were not allowed to make any noises, participants tend to forget that rule over time.
Those social cues are likely to affect inter-brain connections. As they occur randomly they influence each execution of the experiment differently and cannot be standardized. These effects can only even out by having a large amount of participants.

Since only two hyperscanning-related studies are using \ac{vr} \acp{hmd}, it is not well-researched how big the impact of the avatar representation on inter-brain connections is.
Other studies investigated different \ac{fb} avatars, for example, \cite{influence_avatar_representation_this} used a 3D-scanned avatar of the participants which looks closer to a normal human than the abstract model that was used in our experiment.
Yet, it is important to note that, to the best of our knowledge, no other study has combined a comparison of avatars with hyperscanning. So, further research on different avatars and their effects on inter-brain connections needs to be done.
Likewise, it is not clear what influence the conducted task has on the results. In this regard, more investigation needs to be done on how different tasks in \ac{vr} influence inter-brain connections.

\section{Conclusion and Outlook}
\label{sec:conclusion}
In this paper, we have investigated the effects of human avatar representations on inter-brain connections. For this purpose, we have conducted a hyperscanning study with eight pairs. Based on a simple motoric task where the participants had to navigate a ball through a maze, we analyzed different conditions and their implications on inter-brain connectivity. The first condition was done in real life only, so the participants navigated the ball through a custom-made real-physical maze. The other two conditions took place in a \ac{ve}. A room in the \ac{ve} was designed similar to the room the \ac{rl} condition took place in.
Also, the virtual maze was designed exactly like the maze in real life. In the two \ac{vr} conditions, the participants saw each other as a \ac{fb} avatar as well as an \ac{hh} avatar.
During all three conditions, \ac{eeg} data from both participants was measured simultaneously. The main results of our hyperscanning study show that the number of significant sensor pairs was the highest in the \ac{rl} condition, medium in the \ac{fb} VR condition, and lowest in the \ac{hh} condition indicating that an avatar that looks more like a real human enables more significant sensor pairs to appear. Furthermore, when looking at specific sensor pairs that were significant in the majority of participant pairs, the \ac{fb} condition stands out as it shows six inter-brain connections as opposed to only one inter-brain connection during the \ac{hh} condition. However, the results are a first indicator and further steps are necessary to conclude generalizable findings. 

Considering possible future work, the task design, environment design, and avatar representations can influence the effect on inter-brain connections. Therefore, different variants of tasks and environments in various application domains as well as avatar representations can be analyzed to identify their implications on inter-brain connections. Furthermore, knowing more about the physiological implications that inter-brain connections have, could have an impact on brain-computer interfaces. If it is possible for an application to calculate inter-brain connections between its users at runtime, the application could give direct feedback to its users. This could be especially helpful, for example, in a teacher-student situation if the teacher has feedback on how engaged the student is.


\bibliographystyle{splncs04}
\bibliography{references}

\end{document}